\documentclass[sigconf]{acmart}
\usepackage{color}
\usepackage{flushend}
\usepackage{graphicx}
\usepackage{booktabs}
\usepackage[htt]{hyphenat}
\usepackage[]{xcolor}
\usepackage{tabularx}
\usepackage{booktabs}
\usepackage{multirow}
\usepackage{xspace}

\definecolor{midnightgreen}{rgb}{0.0, 0.29, 0.33}

\newcommand{\shortname}{\texttt{iKAT}\xspace}
\newcommand{\fullname}{\texttt{Interactive Knowledge Assistance Track}\xspace}
\newcommand{\RG}{R$\rightarrow$G\xspace}
\newcommand{\GRG}{G$\rightarrow$R$\rightarrow$G\xspace}

\begin{document}
\title[TREC iKAT 2023 Overview]{TREC iKAT 2023: The Interactive Knowledge Assistance Track Overview}

\author{Mohammad Aliannejadi}
\affiliation{%
  \institution{University of Amsterdam}
  \city{Amsterdam}
  \country{The Netherlands}
}
\email{m.aliannejadi@uva.nl}

\author{Zahra Abbasiantaeb}
\affiliation{%
  \institution{University of Amsterdam}
  \city{Amsterdam}
  \country{The Netherlands}
}
\email{z.abbasiantaeb@uva.nl}

\author{Shubham Chatterjee}
\affiliation{%
  \institution{University of Edinburgh}
  \city{Edinburgh}
  \country{Scotland, UK}
}
\email{shubham.chatterjee@ed.ac.uk}

\author{Jeffery Dalton}
\affiliation{%
  \institution{University of Edinburgh}
  \city{Edinburgh}
  \country{Scotland, UK}
}
\email{jeff.dalton@ed.ac.uk}

\author{Leif Azzopardi}
\affiliation{%
  \institution{University of Strathclyde}
  \city{Glasgow}
  \country{Scotland, UK}
}
\email{leif.azzopardi@strath.ac.uk}

\renewcommand{\shortauthors}{Aliannejadi et al.}

\begin{abstract}
    Conversational Information Seeking has evolved rapidly in the last few years with the development of Large Language Models providing the basis for interpreting and responding in a naturalistic manner to user requests.
iKAT emphasizes the creation and research of conversational search agents that adapt responses based on the user's prior interactions and present context. This means that the same question might yield varied answers, contingent on the user’s profile and preferences. The challenge lies in enabling Conversational Search Agents (CSA) to incorporate personalized context to effectively guide users through the relevant information to them. iKAT's first year attracted seven teams and a total of 24 runs. Most of the runs leveraged Large Language Models (LLMs) in their pipelines, with a few focusing on a generate-then-retrieve approach.
\end{abstract}

\maketitle              %
\section{Introduction}

\begin{figure}[t]
    \centering
    \includegraphics[width=\linewidth]{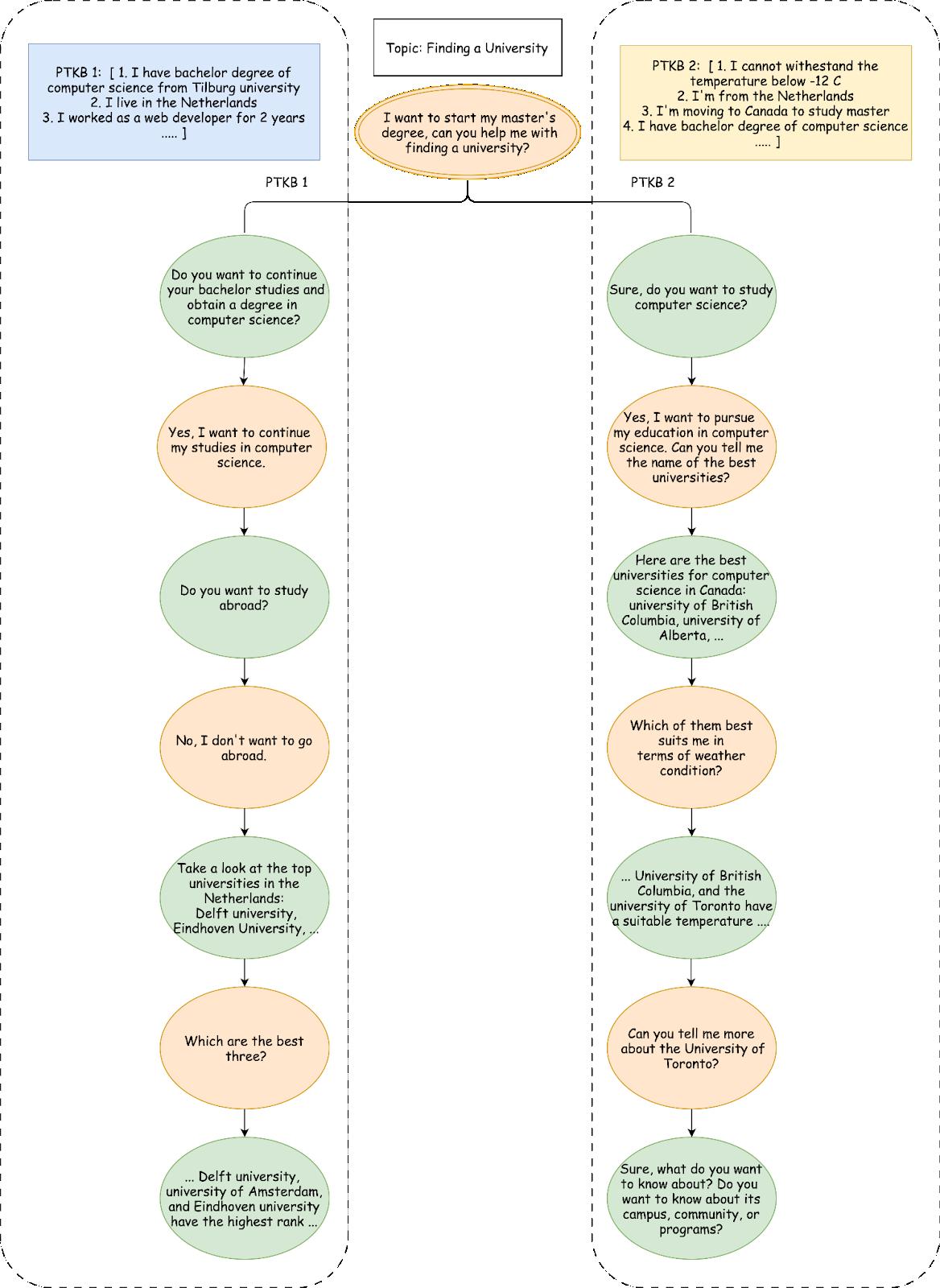}
    \caption{Two flowcharts representing different dialogues between a prospective student and an AI assistant on the topic of finding a suitable university for a master's degree in computer science. On the left, the conversation (PTKB 1) revolves around a student with a bachelor's degree from Tilburg University and work experience, who prefers to stay in the Netherlands. The dialogue suggests top Dutch universities and narrows down to the top three based on ranking. On the right, the second conversation (PTKB 2) involves a student who cannot tolerate cold temperatures below -12°C and is planning to move to Canada for a master's degree. The assistant provides options for top Canadian universities and further refines the suggestions to those with favorable weather conditions, eventually offering detailed information about the University of Toronto upon request. Each conversation flow is guided by the student's preferences, leading to tailored university recommendations.}
    \label{fig:example-intro}
\end{figure}

Conversational Information Seeking stands as a pivotal research area with significant contributions from previous works~\cite{radlinski2017theoretical, azzopardi2018conceptualizing}. The TREC \fullname (\shortname) builds on the foundational work of the TREC Conversational Assistance Track (CAsT)~\cite{owoicho2023trec}. However, \shortname distinctively emphasizes the creation and research of conversational search agents that adapt responses based on user's prior interactions and present context. This means that the same question might yield varied answers, contingent on the user's profile and preferences. Consider a scenario where a user is inquiring about alternatives to cow's milk. Two personas can illustrate this: 
\begin{itemize}
    \item (A) Alice is a vegan who is deeply concerned about the environment.
    \item (B) Bob has been recently diagnosed with diabetes, has a nut allergy, and is lactose intolerant.
\end{itemize}
Given Alice and Bob's personas, their conversation with the system would evolve and develop in very different ways.
This is because what is  relevant to Alice may not necessarily be relevant to Bob, and vice versa. Consequently, by the end of their conversation, what they have learned about, what they have understood, and what they have decided regarding milk alternatives would vary, reflecting their personalized contexts. A detailed concrete example is shown in Figure~\ref{fig:example-intro}.

The challenge lies in enabling Conversational Search Agents (CSA) to incorporate this personalized context to effectively guide users through the relevant information to them. \shortname also emphasizes decisional search tasks~\cite{russell2018information}, where users sift through data and information to weigh up options in order to reach a conclusion or perform an action. These tasks, prevalent in everyday information seeking decisions -- be it related to travel, health, or shopping -- often revolve around subset of high level information operators where queries or questions about the information space include: finding options, comparing options, identifying the pros and cons of options, etc. Given the different personas and their information need (expressed through the sequence of questions), diverse conversation trajectories will arise ---- because the answers to these similar queries will be very different.

In \shortname's debut year, we decided to emphasize these tailored information needs by accounting for a person's knowledge, objectives, tastes, and limitations. To represent their personas, we used a Personal Text Knowledge Base (PTKB) to encapsulate both the task contexts and user specifics. The information requirements encompassed multifaceted tasks, including research, planning, and decision-making processes. Key research questions revolved around:

\begin{enumerate}
    \item \textbf{Personal Contexts}: How efficiently can an agent navigate various personal contexts, leading to distinct, relevant conversations?
    \item \textbf{Personalization}: Can agents adeptly modify conversational feedback based on the user's knowledge?
    \item \textbf{Elicitation}: Are agents proficient in drawing out pertinent persona information to customize discussions?
    \item \textbf{Dependent Relevance}: Can agents effectively employ context and prior responses to foster relevant conversations?
\end{enumerate}

The primary challenge in the track was to deliver a relevant and informative response given the user's PRKB. While these responses could be extractive passages, they might also amalgamate or summarize insights from various passages. Every response, though, should cite at least one ``provenance'' passage from the collection, maintaining a focus on passage/provenance ranking. As in preceding editions of TREC CAsT, systems can leverage all previous conversation turns as context, equivalent to taking the parents in the conversational topic tree.

\section{Track, Tasks, Data, and Resources}
A detailed explanation of the task, data, and resources is provided below. 

\subsection{Track and Tasks}
\label{sec:task}
The focus of the track is on developing a personalized conversational search agent. In our track, the system is provided with some personal information about the user and considers this information in retrieving the relevant documents to the user’s utterance and generating the response. The personal information about the user is provided in the PTKB which is a set of narrative sentences. The sentences are assumed to be collected from previous conversations of the user with the system. Similar to  CAsT~\cite{owoicho2023trec}, the main tasks are passage retrieval and answer generation but considering the persona of the user in understanding the user’s utterance. The track, also, includes the Statement Ranking task where the relevant statements from PTKB to the current user utterance should be identified. To sum up, the track includes the following tasks:

\begin{itemize}
    \item \textbf{Statement Ranking:} The relevant statements from the PTKB for answering the current user utterance should be determined in this step. We approach this task as a relevance score prediction and a ranking task. Given the context of the conversation and user utterance, the system ranks the statements from PTKB based on their relevance and returns a sorted list of PTKB statements. 
    \item \textbf{Passage Ranking:} Given the current user utterance, the context of the conversation, and the PTKB, the system retrieves relevant passages from the collection and ranks them based on their relevance. 
    \item \textbf{Response Generation:} A response is the answer text that is intended to be shown to the user. It should be fluent, satisfy their information need, and not contain extraneous or redundant information. The response could be a generative or abstractive summary of the relevant passages.
\end{itemize}

\subsection{Topics}
\label{sec:topics}
The iKAT 2023 has 11 train and 25 test topics. Each topic includes between 1-3 conversations of different personas on the same topic. 
A personalized turn is defined as a turn that has at least one relevant statement from PTKB. That means the system needs to consider at least one statement in the PTKB in order to answer the user's utterance accurately.

\paragraph{\textbf{Topic creation}} 
A complete set of guidelines was designed for topic creation. The guidelines included a detailed and step-by-step procedure for topic creation and a thorough explanation of the points that should be considered during the process. In addition, the guidelines included a checklist to ensure the quality of the topics. These criteria include both persona-level, turn-level, and conversation-level quality assurance terms.
The topics are generated by organizers and NIST assessors. The topics that did not meet the quality criteria were re-generated by another annotator. Each topic developed was checked and refined by at least two other experts.
The topic creation process included the following steps: 
\begin{enumerate}
    \item generate the user's PTKB for a given conversation/topic;
    \item form the user utterance's for each turn; 
    \item identify the relevant PTKB statements;
    \item retrieve the relevant passages using the searcher tool provided to annotators (iKAT searcher), and; 
    \item form the response of the system.
\end{enumerate}
In generating the PTKB, we took great care to ensure that only high level personal information was included (and any personally identifiable information) was not included to ensure privacy of the contributors.

\subsection{Collection}
\label{sec:collection}
Considering the size of the ClueWeb22-B dataset, we utilized a subset of the ClueWeb22-B collection. 
To create this subset, we manually inspected the domains of the documents within the ClueWeb22-B dataset. We prioritized the diversity of domains and eliminated those that were not relevant. 
The final subset contained 116,838,987 passages, which was distributed by CMU.

To segment the documents into passages, we used a similar methodology as the one used by the TREC Deep Learning track for MS MARCO. We performed the following steps:

\begin{enumerate}
    \item Each document was initially shortened to a length of 10,000 characters.
    \item A sliding window approach was then used, where we took 10 consecutive sentences as a single passage.
    \item After these 10 sentences, we moved the window by 5 sentences (i.e., a 5-sentence stride) to create the next passage.
\end{enumerate}

For the participants, we provided the:
\begin{enumerate}
    \item Python scripts that were used to segment the passages.
    \item Segmented passages along with MD5 hashes.
    \item Pyserini index of the collection.
\end{enumerate}

\subsection{Baselines}
\label{sec:baselines}
The organizers provided four baseline runs detailed below:

\begin{enumerate}
    \item \textbf{bm25\_rm3-manual-ptkb\_3-k\_100-num\_psg-3}. We used BM25+RM3, with the default configuration in Pyserini, to retrieve an initial set of 100 passages for each query. To refine the query, we manually selected the top 3 most relevant PTKB statements and appended them to the manually resolved query. With our rewritten query, we conducted a second round of retrieval using the standard BM25 method in Pyserini. This process also retrieved 100 passages. From this secondary set of 100 passages, we selected the top 3 based on their relevance. These selected passages were then used to construct a final response. For this task, we used a T5 model that has been fine-tuned on the News Summary dataset, available on HuggingFace.\footnote{\url{https://huggingface.co/mrm8488/t5-base-finetuned-summarize-news}}

    \item \textbf{bm25\_rm3-auto-ptkb\_3-k\_100-num\_psg-3}. This approach is analogous to the textit{bm25\_rm3-manual-ptkb\_3-k\_100-num\_psg-3} method but employed an automated processes for query rewriting and PTKB statement selection.  Specifically, we rewrote the query automatically using a T5 model fine-tuned on the Canard dataset,\footnote{\url{https://huggingface.co/castorini/t5-base-canard}} and obtained relevant PTKB statements automatically by re-ranking the statements using \texttt{SentenceTransformers}.\footnote{\url{https://huggingface.co/cross-encoder/ms-marco-MiniLM-L-6-v2}}

    \item \textbf{llama2\_only\_10\_docs}. This pipeline executed several interactions with a LLaMA-2 7B model, each employing distinct prompts tailored for specific tasks. The initial call involved revising the most recent part of the ongoing conversation. The prompt, which included the entire conversation up to that point, was designed to guide the model in reformulating the latest utterance. This step aimed to optimize the utterance for more effective search results in subsequent steps. Following the rewrite, the next step involved evaluating the relevance of documents retrieved based on the revised utterance. In this phase, the prompt fed to the model included both the conversation (as updated from the first call) and a specific document. The model's task was to assess and score the document's relevance in relation to the conversation's context. We only ranked the top 10 documents in the interest of time.  The final call in the pipeline focused on generating an appropriate response to fulfill the user's information need. The prompt for this stage incorporated the top three documents identified as relevant from the previous step, along with the entire conversation. The model uses this information to craft a response that aligned with the user's query and the context provided by the conversation and the selected documents.

    \item \textbf{ColBERT\_llama2summariser\_manual} employed ColBERT for retrieval with manual queries, and LLama-2 7B for summarizing the top-3 passages.

\end{enumerate}

\subsection{PTKB Statement Relevance Assessment}
\label{sec:ptkbRelevance}
To assess the relevance of PTKB statements for each turn, we used two different sets of assessments which were created by the organizers and NIST assessors.

During topic generation, the organizers annotated each turn in terms of their provenance to PTKB statements and included their labels in the released topic files. To ensure the quality of these annotations, we assigned each turn to at least two of the organizers. In cases of disagreement, we assigned the turns to a third annotator and assigned the majority vote label.

Moreover, during the assessment of passage relevance, the NIST assessors were also asked to judge the relevance of PTKB statements to each turn. The assessment pool in this case was smaller than the one done by the organizers. The organizers judged all of the turns, while the NIST assessors only judged the turns that were selected for passage relevance. 
Given such differences, we released and report the performance of the submissions, measured based on both sets of assessments.

\subsection{Passage Retrieval Assessment}
\label{sec:passRet}

\begin{figure}
    \centering
    \includegraphics[width=\columnwidth]{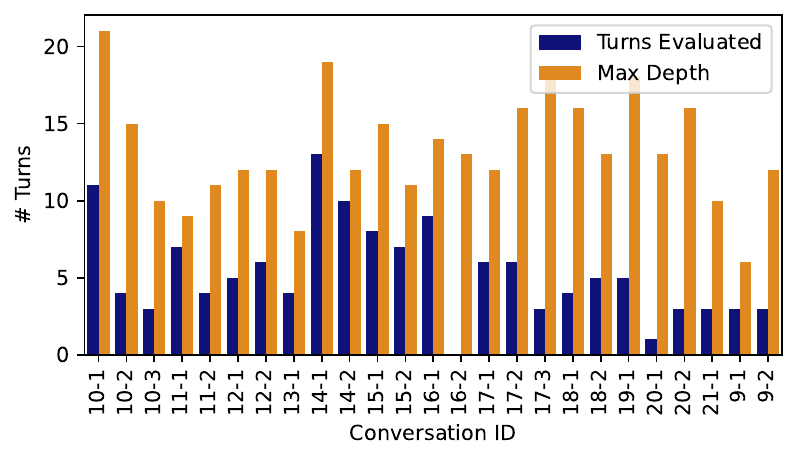}
    \caption{Number turns evaluated per dialogue in the final judgment pool vs.~the maximum depth of each topic.}
    \label{fig:turn-eval}
\end{figure}

The NIST assessors have judged the relevance of the passages based on the methodology used in CAsT (with the same scale). We selected a subset of 176 turns out of 326 to be judged by NIST assessors. Among the un-assessed turns, were responses that were clarifications (e.g., ``Do you have any dietary requirements?'') or were responses to utterances that were too general and returned too many relevant documents (e.g., ``I'm traveling to California, do you have any suggestions?'').
A pool of 26,159 passages was created and manually judged. An average number of 147 passages were judged for each turn. More detailed statistics of the collected data and judgments can be found in Table~\ref{tab:stats}. We also reported the number of turns per dialogue, as well as the number of turns evaluated per dialogue in Figure~\ref{fig:turn-eval}.

\begin{table}[]
    \centering
    \caption{Statistics of test data}
    \begin{tabular}{ll}
    \toprule
        Topics & \phantom{00,0}13 \\
        Dialogues & \phantom{00,0}25 \\
        Turns & \phantom{00,}326 \\
        Assessed turns & \phantom{00,}176 \\ 
        Avg.~dialogue length & \phantom{00,0}13.04 \\
        Avg.~ num.~of dialogue per topic &  \phantom{00,00}1.92\\        
        Passages assessed & 26,159 \\
    \midrule
        Num of pruned turns  & \phantom{00,0}43 \\ 
        Num of turns after pruning  & \phantom{00,}133\\
        Num of dialogues after pruning  & \phantom{00,0}24 \\ 
    \midrule
        Fails to meet (0) & 20,458\\
        Slightly meets (1) & \phantom{0}2,787\\
        Moderately meets (2) & \phantom{0}1,803\\
        Highly meets (3) & \phantom{00,}932\\
        Fully meets (4) & \phantom{00,}179\\
    \midrule \midrule
        PTKB turns assessed by NIST & \phantom{00,0}98 \\
        PTKB assessments by NIST & \phantom{0}1,030 \\
        Relevant (1) & \phantom{00,}224 \\
    \midrule
        PTKB turns assessed by the organizers & \phantom{00,}112 \\
        PTKB assessments by the organizers & \phantom{0}1,158 \\
        Relevant (1) & \phantom{00,}182 \\
    \bottomrule            
    \end{tabular}    
    \label{tab:stats}
\end{table}

\subsection{Response Quality Assessment}
\label{sec:baselines}
An automated approach was taken to assess the quality of responses, where we employed GPT-4. To do this, 
we selected a subset of the turns for assessment, discarding generic turns, while preserving  personalized turns. We assessed the top k responses generated for each turn for each submission. We  also screened the responses and filtered out the low-quality responses. For example, if the response was not semantically similar to the top-ranked passages, or if it included repeated sentences.

Given the subset of turns, we then selected the passages participants indicated that they used to generate the response. In case they did not include the list of used passages, we considered the top 5 passages, as instructed in the guidelines. 
To avoid excessive and unnecessary assessment, we only considered the automatic top run of the teams, in case they were more effective than the median performance. 

Each response was then evaluated from two perspectives: groundedness and naturalness. The criteria and the definitions we provided to GPT-4 for the assessment of each were as follows:

\textbf{Naturalness:} Does the response sound human-like?
\begin{itemize}%
    \item \textit{0. No} - The response does not sound like something a human would say given the conversation.
    \item \textit{1. Somewhat} - Parts of the response can be generated by human, but it is overall not fluent.
    \item \textit{2. Slightly natural} - The response is almost human-like. The response is well-formed but is not natural.
    \item \textit{3. Yes (but not completely)} - The response is almost human-like. The response is well-formed and natural in most parts but has some parts that are not natural.
    \item \textit{3. Yes} - The response is perfectly human-like and fluent.
\end{itemize}

\textbf{Groundedness:} Does the response appropriately reference or connect to the information provided in the provenance passages? 

\begin{itemize}%
    \item \textit{0: No} - The response does not reference the information provided in the provenance passages or is entirely disconnected from it.
    \item \textit{1: Somewhat} - The response contains elements that hint at the provenance passages but lacks a clear connection or misinterprets the information.
    \item \textit{2: Yes (but not completely)} - The response reflects the information provided in the provenance passages but may not fully utilize all relevant details or may include minor inaccuracies.
    \item \textit{3: Yes} - The response is directly based on the information provided in the provenance passages, accurately reflects this information, and utilizes it to enhance the response's relevance and completeness.
\end{itemize}

For naturalness, only the response was provided with the instructions, while for groudness, the response and the provenance passages were included.
To ensure the quality of the assessments, we tested multiple setups and prompts and compared them to a subset of responses that were manually labeled by the organizers. We used the setting that had the highest agreement with the labeled data.

\section{Evaluation}
\label{sec:eval}

\paragraph{Statement Ranking Task} We evaluated the PTKB statement ranking task at the turn and conversation levels. The ranking metrics include nDCG@3, P@3, Recall@5, and MRR.

\paragraph{Passage Ranking Task.} {For the main task, we evaluated the runs across two dimensions given the ranking for each topic turn: (i) the ranking depth and (ii) the turn depth. For ranking depth, we focused on earlier positions 3 and 5 for the conversational scenario (where we assumed that the top $k$ results would be used to formulate the response). 
Then for turn depth we evaluated the run performance at the n-th conversational turn. Performing well on deeper rounds indicates a better ability to understand the preceding context. 
We used the mean nDCG@5 as the main evaluation metric, with all conversational turns averaged using uniform weights. We also measured the turned-depth measure based on nDCG@5, with the per turn nDCG@5 scores averaged at depth ($n$). In addition to the nDCG metrics (nDCG, nDCG@3, and nDCG@5), we also calculated P@10, Recall, Recall@10, and mean Average Precision, where again, we averaged over all turns.}

\paragraph{Response Generation Task.} Given the high likelihood of LLMs being used in this year's submissions and the possibility of hallucination, we evaluated the generated responses in terms of groundedness. Groundedness measures whether the generated response can be attributed to the passages that it is supposed to be generated from. We use GPT-4 to evaluate both the groundedness and naturalness of the responses, as it demonstrated a high correlation with human labels in our preliminary experiments. For each turn we used the GPT-4 assessments, and took the mean of groundedness and naturalness.
over all turns.

\begin{table*}[t]
    \centering
    \caption{Participants and their runs.}
    \label{tab:participants}
    \small
    \begin{tabular}{llll}
\toprule
           Group &                                            Run ID &        run\_type & Pipeline \\
\midrule
       CFDA\_CLIP &                                             cfda4 &       automatic &     R$\rightarrow$G \\
       CFDA\_CLIP &                                             cfda3 &       automatic &     R$\rightarrow$G \\
       CFDA\_CLIP &                                             cfda1 &       automatic &     R$\rightarrow$G \\
       CFDA\_CLIP &                                             cfda2 &       automatic &     R$\rightarrow$G \\
      GRILL\_Team &  GRILL\_BM25\_T5Rewriter\_T5Ranker\_BARTSummariser\_10 &       automatic &     R$\rightarrow$G \\
      GRILL\_Team &                     GRILL\_Colbert\_BART2Summariser &       automatic &     R$\rightarrow$G \\
      GRILL\_Team &     GRILL\_BM25\_T5Rewriter\_T5Ranker\_BARTSummariser &       automatic &     R$\rightarrow$G \\
            IITD &                        run\_automatic\_dense\_monot5 &       automatic &     R$\rightarrow$G \\
            IITD &              run\_automatic\_dense\_mini\_LM\_reranker &       automatic &     R$\rightarrow$G \\
            IITD &                            run\_automatic\_llm\_damo &       automatic &     R$\rightarrow$G \\
            IITD &      run\_automatic\_dense\_damo\_canard\_16000\_recall &       automatic &     R$\rightarrow$G \\
 IRLab-Amsterdam &                             run-1-llama-zero-shot &       automatic &     R$\rightarrow$G \\
 IRLab-Amsterdam &                            run-2-llama-fine-tuned &       automatic &     R$\rightarrow$G \\
 IRLab-Amsterdam &                                       run-4-GPT-4 &       automatic &  G$\rightarrow$R$\rightarrow$G \\
 IRLab-Amsterdam &                     run-3-llama-fine-tuned-manual &     manual-both &     R$\rightarrow$G \\
       InfoSense &                   georgetown\_infosense\_ikat\_run\_2 &       automatic &  G$\rightarrow$R$\rightarrow$G \\
       InfoSense &                   georgetown\_infosense\_ikat\_run\_3 &       automatic &  G$\rightarrow$R$\rightarrow$G \\
       InfoSense &                   georgetown\_infosense\_ikat\_run\_1 &       automatic &  G$\rightarrow$R$\rightarrow$G \\
      Organizers &              bm25\_rm3-auto-ptkb\_3-k\_100-num\_psg-3 &       automatic &     R$\rightarrow$G \\
      Organizers &            bm25\_rm3-manual-ptkb\_3-k\_100-num\_psg-3 &     manual-both &     R$\rightarrow$G \\
      Organizers &                   colbert\_llama2summariser\_manual &  manual-rewrite &     R$\rightarrow$G \\
      Organizers &                               llama2\_only\_10\_docs &       automatic &     R$\rightarrow$G \\
            RALI &                                           ConvGQR &       automatic &     R$\rightarrow$G \\
            RALI &                                        LLMConvGQR &       automatic &     R$\rightarrow$G \\
       uot-yj &                              uot-yj\_run\_monot5 &       automatic &     R$\rightarrow$G \\
       uot-yj &                           uot-yj\_run\_rankgpt35 &       automatic &     R$\rightarrow$G \\
       uot-yj &                           uot-yj\_run\_llmnoptkb &       automatic &     R$\rightarrow$G \\
       uot-yj &                                     uot-yj\_run &       automatic &     R$\rightarrow$G \\
\bottomrule
\end{tabular}

\end{table*}

\begin{table*}[t]
\centering
\caption{Automatic evaluation of passage retrieval results. \GRG run names are highlighted with \textit{italic} font. Evaluation at retrieval cutoff of 1000.
}
 \label{tab:automatic-results}
  \resizebox{\textwidth}{!}{%
    \begin{tabular}{llrrrrrrr}
\toprule
          Group &                                           Run ID &  nDCG@3 &  nDCG@5 &   nDCG &   P@20 &  Recall@20 &  Recall &    mAP \\
\midrule
IRLab-Amsterdam &                                      \textit{run-4-GPT-4} &  0.4382 &  0.4396 & 0.3479 & 0.3444 &     0.1821 &  0.3456 & 0.1759 \\     
      InfoSense &                  \textit{georgetown\_infosense\_ikat\_run\_3} &  0.3083 &  0.3109 & 0.2097 & 0.2519 &     0.1168 &  0.1862 & 0.1042 \\
      InfoSense &                  \textit{georgetown\_infosense\_ikat\_run\_2} &  0.2912 &  0.2955 & 0.2119 & 0.2643 &     0.1211 &  0.1862 & 0.1072 \\
      InfoSense &                  \textit{georgetown\_infosense\_ikat\_run\_1} &  0.2292 &  0.2299 & 0.1689 & 0.2109 &     0.1015 &  0.1613 & 0.0868 \\
           IITD &                       run\_automatic\_dense\_monot5 &  0.2167 &  0.2206 & 0.2147 & 0.1831 &     0.0812 &  0.3058 & 0.0754 \\
           RALI &                                          ConvGQR &  0.1652 &  0.1623 & 0.1518 & 0.1421 &     0.0611 &  0.2034 & 0.0551 \\
           IITD &     run\_automatic\_dense\_damo\_canard\_16000\_recall &  0.1648 &  0.1619 & 0.1352 & 0.1402 &     0.0557 &  0.1664 & 0.0505 \\
      uot-yj &                          uot-yj\_run\_llmnoptkb &  0.1433 &  0.1469 & 0.0759 & 0.1071 &     0.0525 &  0.0525 & 0.0350 \\
     Organizers &                              llama2\_only\_10\_docs &  0.1389 &  0.1466 & 0.0756 & 0.1192 &     0.0553 &  0.0553 & 0.0376 \\
IRLab-Amsterdam &                            run-1-llama-zero-shot &  0.1494 &  0.1437 & 0.0815 & 0.1165 &     0.0507 &  0.0742 & 0.0387 \\
           IITD &                           run\_automatic\_llm\_damo &  0.1343 &  0.1411 & 0.1105 & 0.1102 &     0.0487 &  0.1401 & 0.0376 \\
           RALI &                                       LLMConvGQR &  0.1318 &  0.1338 & 0.1200 & 0.1169 &     0.0523 &  0.1620 & 0.0461 \\
      CFDA\_CLIP &                                            cfda1 &  0.1323 &  0.1291 & 0.0941 & 0.1267 &     0.0536 &  0.0963 & 0.0444 \\
      CFDA\_CLIP &                                            cfda2 &  0.1282 &  0.1260 & 0.0916 & 0.1218 &     0.0510 &  0.0963 & 0.0421 \\
      uot-yj &                          uot-yj\_run\_rankgpt35 &  0.1130 &  0.1070 & 0.0496 & 0.0801 &     0.0322 &  0.0322 & 0.0224 \\
      uot-yj &                             uot-yj\_run\_monot5 &  0.1107 &  0.1062 & 0.0499 & 0.0823 &     0.0330 &  0.0330 & 0.0223 \\
      uot-yj &                                    uot-yj\_run &  0.1086 &  0.1049 & 0.0495 & 0.0823 &     0.0330 &  0.0330 & 0.0222 \\
           IITD &             run\_automatic\_dense\_mini\_LM\_reranker &  0.1056 &  0.1047 & 0.0548 & 0.0812 &     0.0308 &  0.0496 & 0.0206 \\
IRLab-Amsterdam &                           run-2-llama-fine-tuned &  0.0826 &  0.0816 & 0.0457 & 0.0684 &     0.0301 &  0.0425 & 0.0202 \\
      CFDA\_CLIP &                                            cfda4 &  0.0836 &  0.0806 & 0.0759 & 0.0793 &     0.0362 &  0.0963 & 0.0311 \\
      CFDA\_CLIP &                                            cfda3 &  0.0836 &  0.0806 & 0.0759 & 0.0793 &     0.0362 &  0.0963 & 0.0311 \\     
     GRILL\_Team &                    GRILL\_Colbert\_BART2Summariser &  0.0667 &  0.0641 & 0.0558 & 0.0451 &     0.0278 &  0.0669 & 0.0184 \\
     GRILL\_Team &    GRILL\_BM25\_T5Rewriter\_T5Ranker\_BARTSummariser &  0.0630 &  0.0620 & 0.0496 & 0.0579 &     0.0214 &  0.0636 & 0.0168 \\
     GRILL\_Team & GRILL\_BM25\_T5Rewriter\_T5Ranker\_BARTSummariser\_10 &  0.0572 &  0.0581 & 0.0356 & 0.0500 &     0.0224 &  0.0284 & 0.0172 \\
     Organizers &             bm25\_rm3-auto-ptkb\_3-k\_100-num\_psg-3 &  0.0396 &  0.0450 & 0.0277 & 0.0429 &     0.0176 &  0.0257 & 0.0118 \\
\bottomrule
\end{tabular}

}

\end{table*}

\begin{figure*}
    \centering
    \includegraphics[width=0.6\textwidth]{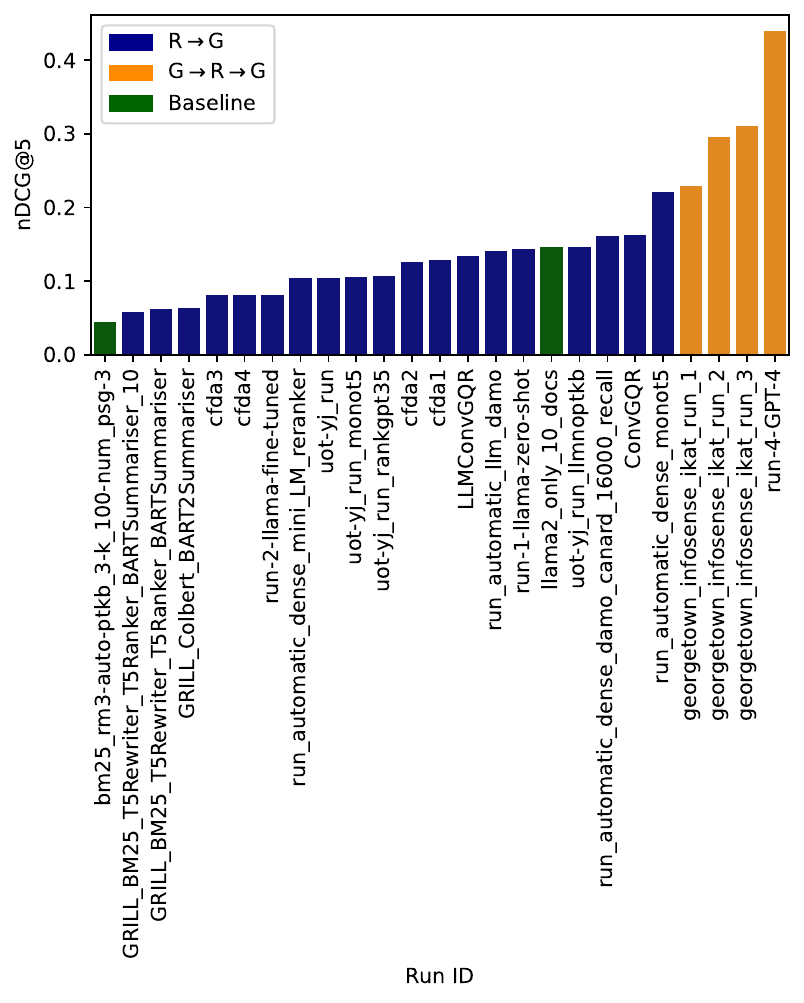}
    \caption{Performance of all automatic runs in terms of nDCG@5 on the passage ranking task. }
    \label{fig:automaitc_runs}
\end{figure*}

\begin{table*}[t]
\centering
\caption{Automatic evaluation of passage retrieval results. 
Evaluation at retrieval cutoff of 1000.
}
 \label{tab:manual-results}
  \resizebox{\textwidth}{!}{%
    \begin{tabular}{llrrrrrrr}
    \toprule
               Group &                                              Run ID &  nDCG@3 &  nDCG@5 &   nDCG &   P@20 &  Recall@20 &  Recall &    mAP \\
    \midrule    
    IRLab-Amsterdam &                    run-3-llama-fine-tuned-manual &  0.4122 &  0.4264 & 0.3245 & 0.3530 &     0.2063 &  0.3157 & 0.1911 \\
         Organizers &           bm25\_rm3-manual-ptkb\_3-k\_100-num\_psg-3 &  0.3319 &  0.3291 & 0.2533 & 0.2767 &     0.1518 &  0.2564 & 0.1349 \\         
        Organizers &            colbert\_llama2summariser\_manual &  0.0693 &  0.0669 & 0.0575 & 0.0474 &     0.0285 &  0.0683 & 0.0191 \\
         
    \bottomrule
    \end{tabular}
}

\end{table*}

\section{Participants}

The iKAT main task received 24 run submissions from seven groups shown in~Table~\ref{tab:participants}. The organizers provided four runs (two automatic, two manual) as baselines for comparison. 
Participants provided metadata and descriptions of their runs.

Most of the runs used LLMs in their pipelines, and we observed two main pipelines, namely, retrieve then generate (\RG) and generate, retrieve, then generate (\GRG).
Most teams used a multi-step \RG pipeline consisting of the following: 
(1) PTKB statement relevance prediction;
(2) conversational rewriting (most incorporating the previous canonical responses as well as predicted relevance PTKB statements) and conversational query expansion;
(3) retrieval using traditional or dense IR model; and 
(4) multi-stage passage re-ranking with neural language models fine-tuned for point-wise (mono) and pairwise (duo) ranking. 
Table~\ref{tab:participants} lists the submissions from the teams, as well as their pipelines.

\subsection{Participant Runs}
\label{subsec:Participant Runs}

Table \ref{tab:participants} provides an overview of the participant runs, and below we include a summary of each:

\begin{enumerate}

\item \textbf{uot-yj\_run}. Pyserini's \texttt{LuceneSearcher} is utilized for initial passage retrieval at each utterance turn. Subsequently, multiple Large Language Models (LLMs) are employed to re-rank the top five passages retrieved during each turn using pair-wise ranking. The results from these LLMs are aggregated to form a final ranking. Both stages, passage retrieval and re-ranking, take into account the first two relevant PTKB statements generated by automated runs, adhering to a zero-shot learning approach.

\item \textbf{uot-yj\_run\_llmnoptkb}. Passage retrieval for each utterance turn is conducted using Pyserini's \texttt{LuceneSearcher}. For re-ranking, multiple LLMs, including \texttt{stabilityai/stablelm-tuned-alpha-7b}, \texttt{eachadea/vicuna-13b-1.1}, \texttt{jondurbin/airoboros-7b}, and \texttt{TheBloke/koala-13B-HF}, are used to re-rank the top five passages retrieved in each turn by pair-wise ranking. An aggregation of these results leads to a final ranking. Notably, both passage retrieval and re-ranking stages do not consider relevant PTKB statements and rely solely on rewritten utterances in each turn. This process also follows a zero-shot learning approach.

\item \textbf{uot-yj\_run\_rankgpt35}. In this approach, Pyserini's \texttt{LuceneSearcher} is used for passage retrieval at each utterance turn. The re-ranking process is conducted using a combination of the GPT-3.5 model with prompts and a sliding window technique. Both the retrieval and re-ranking stages include consideration of the top-2 relevant PTKB statements from automatic PTKB statement ranking runs in each turn, aligning with a zero-shot learning approach.

\item \textbf{uot-yj\_run\_monot5}. Passage retrieval for each utterance turn is performed using Pyserini's \texttt{LuceneSearcher}. The MonoT5 model is then employed for the passage re-ranking process. Similar to the previous methods, both stages, passage retrieval and re-ranking, consider the top-2 relevant PTKB statements derived from automatic PTKB statement ranking in each turn. This method also is based on a zero-shot learning approach.

\item \textbf{georgetown\_infosense\_ikat\_run\_1}. This method first uses the LLaMA model to generate a response to the user's query, which includes PTKBs determined to be relevant. This initial response is then analyzed to identify reliable passages using a TF-IDF and logarithmic regression model. The LLaMa model processes these reliable passages to generate a secondary response. Finally, each passage is summarized into one or two sentences using the FastChat T5 model. These summaries are ranked by relevance to the query and combined to form the final response text.

\item \textbf{georgetown\_infosense\_ikat\_run\_2}. This approach also begins with the LLaMA model generating a response to the user's utterance, integrating relevant PTKBs. The response is then used to find passages classified as reliable by the TF-IDF and logarithmic regression model. Unlike the two-shot approach, passages are directly summarized using FastChat T5, creating one or two-sentence summaries for each. These are ranked by relevance to the query and combined, forming the final text provided as a response.

\item \textbf{georgetown\_infosense\_ikat\_run\_3}. Similar to run 2, this one-shot approach uses the LLaMA model to respond to the user's utterance, incorporating automatically determined relevant PTKBs. The difference lies in the selection of top passages, which are identified based on BM25 scoring instead of the TF-IDF and logarithmic regression model. Each selected passage is then succinctly summarized into one or two sentences using FastChat T5. These sentence summaries, ordered by their relevance to the query, are then combined to create the final response text.

\item \textbf{GRILL\_BM25\_T5Rewriter\_T5Ranker\_BARTSummariser}. Involves T5 for query rewriting and document/passage reranking, BM25 for initial retrieval, and BART for response generation. A LLaMA-based simulator provides simulated user feedback and answers clarification questions.

\item \textbf{GRILL\_Colbert\_BART2Summariser}. Uses ColBERT for dense retrieval and generates responses from the top-3 passages. Includes a LLaMA-2 based simulator for providing feedback and answering clarifying questions.

\item\textbf{GRILL\_BM25\_T5Rewriter\_T5Ranker\_BARTSummariser\_10}. Utilizes T5 for query rewriting, BM25 for initial retrieval, T5 for passage ranking, and LLaMA-2 based simulator for up to 10 rounds per query.

\item \textbf{ConvGQR}. Combines query rewriting and query expansion to train on the QReCC dataset, then applied to the iKAT dataset.

\item \textbf{LLMConvGQR}. Merges query rewriting and query expansion based on ChatGPT, applying query reformulation directly on the iKAT dataset.

\item \textbf{run\_automatic\_dense\_mini\_LM\_reranker}.Two key aspects define this method. Firstly, a neural model is utilized to preserve essential elements during reformulation, ensuring that key items in the conversation are maintained. Secondly, the approach follows conventional dense retrieval, which is then complemented by neural re-ranking.

\item \textbf{run\_automatic\_llm\_damo}. This method operates as a two-step pipeline. The initial step involves dense retrieval of passages, succeeded by re-ranking. Automatic queries are rewritten using a custom-trained query rewriting module, which is based on BART and fine-tuned on the Samsum and Canard datasets. The re-ranking process is executed using the COROM model.

\item \textbf{run\_automatic\_dense\_monot5}. The process here also unfolds in two distinct steps. The first step encompasses dense retrieval of passages, followed by their re-ranking. In this method, automatic queries undergo rewriting through a custom-trained module based on BART, with fine-tuning conducted using the Samsum and Canard datasets. The re-ranking phase employs a T5-based model.

\begin{figure*}
    \centering
    \includegraphics[width=\textwidth]{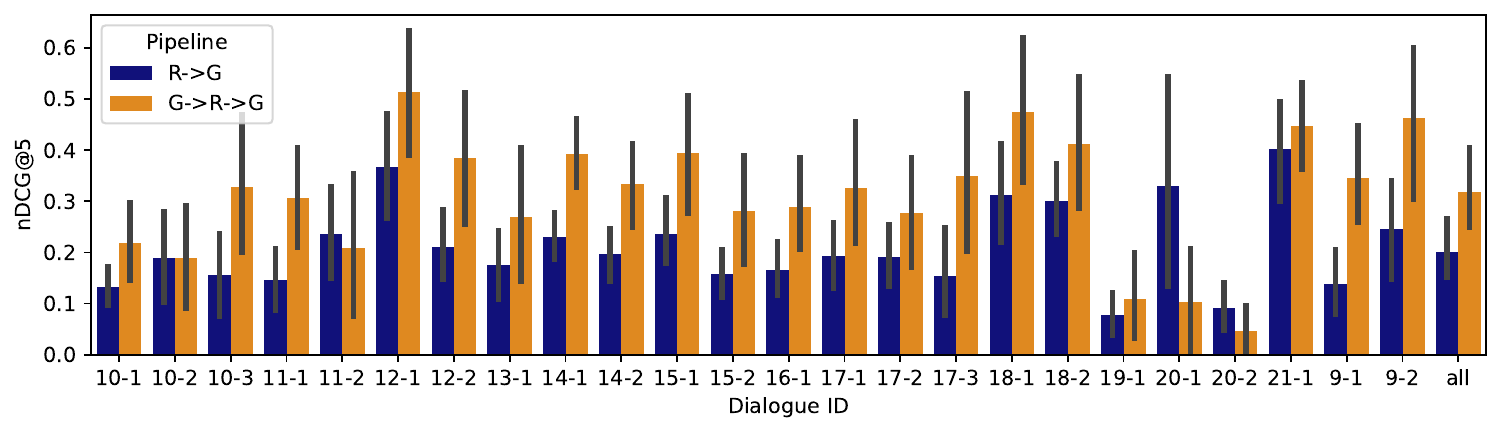}
    \caption{nDCG@5 aggregated for each topic across all runs on the passage ranking task. We report the average across runs, median or better.}
    \label{fig:dialogue-performance}
\end{figure*}

\item \textbf{run\_automatic\_dense\_damo\_canard\_16000\_recall}. This method is structured as a two-step pipeline. Initially, there is dense retrieval of passages, which is then followed by a re-ranking phase. For query rewriting, a module based on T5 is used. The re-ranking of passages is carried out using a T5 based model, aligning with the method's overall structure.

\item \textbf{run-4-GPT-4}. In this run, the GPT-4 model initially generates an answer for each turn. Subsequently, GPT-4 is employed to produce five queries for each answer. These generated queries are then processed by a BM25 model and a cross-encoder \texttt{MiniLM} for re-ranking. The first two documents retrieved for each query are selected and supplied to GPT-4, which then generates the response text.

\item \textbf{run-2-llama-fine-tuned}. For this task, the LLaMa model undergoes fine-tuning, specifically for query rewriting and response generation. \texttt{SentenceTransformers} are used for PTKB selection, and a \texttt{MiniLM12} cross-encoder from HuggingFace is employed for re-ranking.

\item \textbf{run-1-llama-zero-shot}. Query understanding and response generation in this run are based on zero-shot prompting of the LLaMa model, with no training data used for these tasks. Re-ranking is conducted using the cross-encoder model from HuggingFace, specifically the \texttt{ms-marco-MiniLM-L-12-v2} model, which is trained for passage ranking on the MS Marco dataset. This approach applies zero-shot prompting with the LLaMa 7b model for both response generation and query rewriting. \texttt{SentenceTransformers} is utilized for PTKB selection. For re-ranking, the cross-encoder model from HuggingFace (\texttt{ms-marco-MiniLM-L-12-v2}), trained on the MS Marco dataset, is used.

\item \textbf{run-3-llama-fine-tuned-manual}. In this manual run, queries are rewritten manually. Re-ranking is performed using the cross-encoder model from HuggingFace, specifically the \texttt{ms-marco-MiniLM-L-12-v2} model, which is trained for passage ranking on the MS Marco dataset. The LLaMa 7b model, fine-tuned on the iKAT training dataset, is employed for generating responses. The manually rewritten queries and relevant ground truth PTKB statements are re-ranked using BM25.

\item \textbf{cfda1}. The datasets used include QReCC and CAsT. Dense retrieval model trained on the MS MARCO passage ranking collection. The retrieval process involves sparse retrieval, where re-ranking is performed by dense retrievers. Generative QA models are used for response generation.

\item \textbf{cfda2}. The QReCC, CAsT, and MSMARCO passage ranking datasest are utilized. The retrieval process includes sparse retrieval with re-ranking carried out by dense retrievers. For generating responses, generative QA models are employed.

\item \textbf{cfda3}. This method utilizes the QReCC and CAsT datasets. Dense retrieval models trained on the MS MARCO passage ranking collection. The query rewriting process is carried out by a model conditioned on all provided PTKBs. Sparse retrieval is followed by re-ranking executed by dense retrievers, which have been fine-tuned with synthetic statements from QReCC. The response generation employs generative QA models that are fine-tuned on an augmented QReCC dataset with synthetic statements.

\item \textbf{cfda4}. In this method, QReCC, CAsT, and MSMARCO passage ranking datasets are used. Query rewriting is conducted using a statement-aware QR model. The retrieval process involves sparse retrieval with subsequent re-ranking performed by dense retrievers. Response generation is done through generative QA models.
\end{enumerate}

\begin{figure}
    \centering
    \includegraphics[width=\columnwidth]{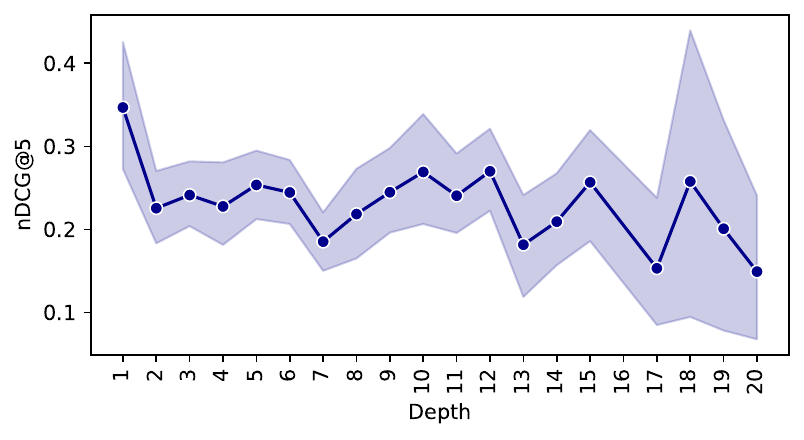}
    \caption{nDCG@5 at varying conversation turn depths on the passage ranking task. We report the average across runs, median or better.}
    \label{fig:turn-performance}
\end{figure}

\begin{figure}
    \centering
    \includegraphics[width=\columnwidth]{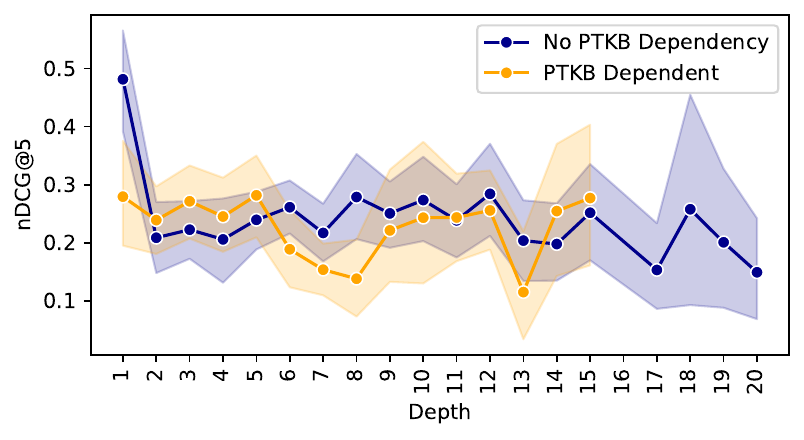}
    \caption{nDCG@5 at varying conversation turn depths on the passage ranking task, for turns that depend on PTKB statements vs.~those that do not. We report the average across runs, median or better.}
    \label{fig:turn-performance-ptkb}
\end{figure}

\section{Results}

\subsection{Passage Ranking}

\subsubsection{Overall results.}
Table~\ref{tab:automatic-results} lists the performance of the automatic runs in terms of all the evaluation metrics. We see that \GRG runs tend to perform better than \RG models, suggesting that leveraging the learned knowledge of LLMs (GPT-4 and Llama in this case) leads to a better starting point for subsequent retrieval of relevant results and then the generation of a relevant response.
Figure~\ref{fig:automaitc_runs} compares the performance of all the automatic runs in terms of nDCG@5, where the runs are color-coded based on their pipelines.
We received three manual runs this year, listed in Table~\ref{tab:manual-results}, where all the submissions follow the \RG pipeline.

\subsubsection{Performance per dialogue.} 
Figure~\ref{fig:dialogue-performance} reports the average performance in terms of nDCG@5 of all runs that median or better. 
We see that while the runs perform well for some of the topics, they fail to perform well for some. In particular, we find topics 12-1 and 21-1 to be the easiest, while 19-1 and 20-2 to be the most difficult.

\subsubsection{Performance at different depths.} 
Figure~\ref{fig:turn-performance} reports the performance of all runs (median or better) at varying conversation turns in terms of nDCG@5. We also report the performance at different depths, separating the turns that depend on PTKB provenance in Figure~\ref{fig:turn-performance-ptkb}. Our intuition is that PTKB statement ranking step will introduce additional difficulty and error in the pipeline and consequently the runs exhibit lower performance. However, we see that this was not always the case, and in most cases PTKB dependence led to lower performance. 
Similar to CAsT, we see that the models perform best in the first turn, and as the conversation progresses the performance becomes lower, with some peaks in the middle of the conversation.
As we compare the performance of the turns based on PTKB dependence, interestingly we see the highest difference in the first turn, suggesting that the significance of predicting the right PTKB statements in the early turns is essential, and the task is more difficult in those earlier turns.

\begin{table*}
\caption{Performance of automatic runs on the PTKB provenance task based on NIST assessment. \GRG run names are highlighted with \textit{italic} font.}
 \label{tab:ptkb-results}
    \begin{tabular}{llrrrr}
\toprule
          Group &                                           Run ID &  nDCG@3 &    P@3 &  Recall@3 &    MRR \\
\midrule
IRLab-Amsterdam &                            run-1-llama-zero-shot &  0.7254 & 0.4626 &    0.6964 & 0.7950 \\
IRLab-Amsterdam &                           run-2-llama-fine-tuned &  0.7102 & 0.4490 &    0.6796 & 0.7795 \\
      uot-yj &                                    uot-yj\_run &  0.6594 & 0.4184 &    0.6213 & 0.7112 \\
IRLab-Amsterdam &                                      \textit{run-4-GPT-4} &  0.6174 & 0.3605 &    0.5833 & 0.7027 \\
      InfoSense &                  \textit{georgetown\_infosense\_ikat\_run\_3} &  0.4515 & 0.2551 &    0.4133 & 0.5446 \\
      InfoSense &                  \textit{georgetown\_infosense\_ikat\_run\_2} &  0.4515 & 0.2551 &    0.4133 & 0.5446 \\
      InfoSense &                  \textit{georgetown\_infosense\_ikat\_run\_1} &  0.4515 & 0.2551 &    0.4133 & 0.5446 \\
     GRILL\_Team &                    GRILL\_Colbert\_BART2Summariser &  0.3727 & 0.2483 &    0.3836 & 0.5038 \\
     Organizers &             bm25\_rm3-auto-ptkb\_3-k\_100-num\_psg-3 &  0.3434 & 0.2687 &    0.3099 & 0.3844 \\
           RALI &                                          ConvGQR &  0.2934 & 0.2109 &    0.2756 & 0.4419 \\
           RALI &                                       LLMConvGQR &  0.2934 & 0.2109 &    0.2756 & 0.4419 \\
     GRILL\_Team & GRILL\_BM25\_T5Rewriter\_T5Ranker\_BARTSummariser\_10 &  0.2605 & 0.2211 &    0.2964 & 0.3757 \\
     GRILL\_Team &    GRILL\_BM25\_T5Rewriter\_T5Ranker\_BARTSummariser &  0.2507 & 0.2075 &    0.3016 & 0.3756 \\
\bottomrule
\end{tabular}

\end{table*}

\begin{table*}
\caption{Performance of automatic runs on the PTKB provenance task based on the organizers' assessment. \GRG run names are highlighted with \textit{italic} font.}
 \label{tab:ptkb-results-org}
    \begin{tabular}{llrrrr}
\toprule
          Group &                                           Run ID &  nDCG@3 &    P@3 &  Recall@3 &    MRR \\
\midrule
IRLab-Amsterdam &                            run-1-llama-zero-shot &  0.6394 & 0.3810 &    0.7375 & 0.6707 \\
      uot-yj &                                    uot-yj\_run &  0.6370 & 0.3512 &    0.6903 & 0.6890 \\
IRLab-Amsterdam &                                      \textit{run-4-GPT-4} &  0.6288 & 0.3423 &    0.6888 & 0.6618 \\
IRLab-Amsterdam &                           run-2-llama-fine-tuned &  0.6149 & 0.3542 &    0.6918 & 0.6617 \\
      InfoSense &                  \textit{georgetown\_infosense\_ikat\_run\_3} &  0.4146 & 0.2232 &    0.4552 & 0.4449 \\
      InfoSense &                  \textit{georgetown\_infosense\_ikat\_run\_2} &  0.4146 & 0.2232 &    0.4552 & 0.4449 \\
      InfoSense &                  \textit{georgetown\_infosense\_ikat\_run\_1} &  0.4146 & 0.2232 &    0.4552 & 0.4449 \\
     Organizers &             bm25\_rm3-auto-ptkb\_3-k\_100-num\_psg-3 &  0.3200 & 0.1905 &    0.3720 & 0.3438 \\
     GRILL\_Team &    GRILL\_BM25\_T5Rewriter\_T5Ranker\_BARTSummariser &  0.2635 & 0.1696 &    0.3052 & 0.3905 \\
     GRILL\_Team &                    GRILL\_Colbert\_BART2Summariser &  0.2457 & 0.1756 &    0.2967 & 0.3659 \\
           RALI &                                          ConvGQR &  0.2227 & 0.1548 &    0.2777 & 0.3442 \\
           RALI &                                       LLMConvGQR &  0.2227 & 0.1548 &    0.2777 & 0.3442 \\
     GRILL\_Team & GRILL\_BM25\_T5Rewriter\_T5Ranker\_BARTSummariser\_10 &  0.2112 & 0.1577 &    0.2640 & 0.3360 \\
\bottomrule
\end{tabular}

\end{table*}

\subsection{PTKB Provenance}

\subsubsection{Overall results.} 
As previously described, we evaluated the submissions for the PTKB statement ranking task based on two relevance judgments, namely, assessed by the NIST assessors, as well as the organizers. We report the results based on NIST assessments in Table~\ref{tab:ptkb-results}, and the results based on the organizers' assessment in Table~\ref{tab:ptkb-results-org} in terms of all evaluation metrics. We see a high agreement between the two tables in the relative order of the submissions. It is worth noting that in both cases, we see that \GRG models are not the top runs, despite their success in passage ranking, suggesting that while the LLMs can leverage PTKB statements effectively in response generation, they are not as effective in ranking the relevant PTKB statements in the \GRG pipeline. 
Llama in the zero-shot setting, however, achieved the best result in PTKB statement ranking task based on both results.

\begin{figure}
    \centering
    \includegraphics[width=\columnwidth]{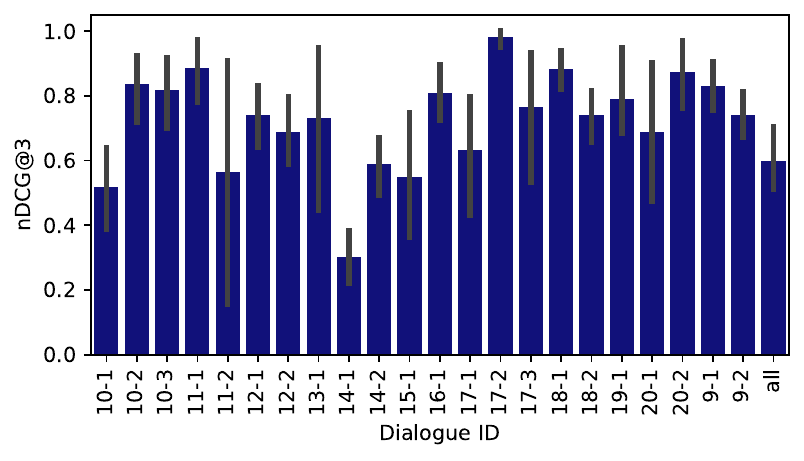}
    \caption{nDCG@3 on PTKB relevance prediction, aggregated for each topic across all runs. We report the average across runs, median or better.}
    \label{fig:ptkb-dialogue}
\end{figure}

\begin{figure}
    \centering
    \includegraphics[width=\columnwidth]{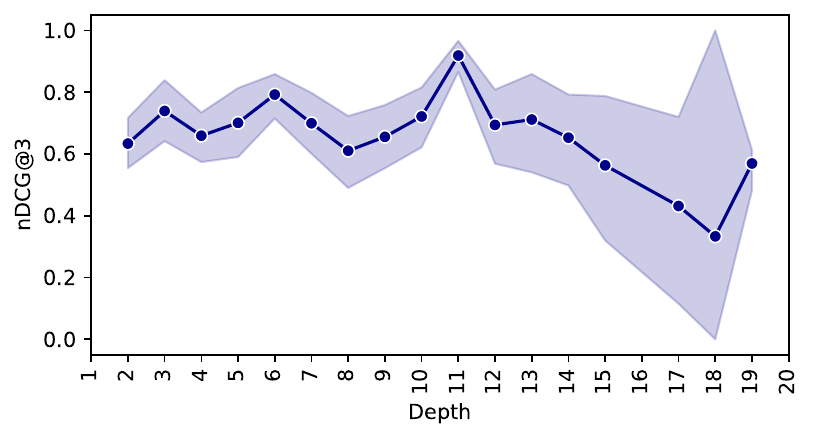}
    \caption{nDCG@3 on PTKB relevance prediction at varying conversation turn depths. We report the average across runs, median or better.}
    \label{fig:ptkb-turn}
\end{figure}

\subsubsection{Performance per dialogue.}
Using the organizers' assessments, in Figure~\ref{fig:ptkb-dialogue} we plotted the mean performance of all the submissions (median and better) in terms of nDCG@3, aggregated on each topic. While we observed a reasonably high performance for all the topics, we find topic 20-2 to be the most challenging for this task, and 16-1 to be among the easiest one. Comparing the results of this table with Table~\ref{fig:dialogue-performance}, surprisingly we do not notice a clear correlation between PTKB statement ranking and passage retrieval performance.

\subsubsection{Performance at different depths.} 
Using the organizers' assessments, in Figure~\ref{fig:ptkb-turn} we plot the mean performance of all the submissions (median and better) in terms of nDCG@3, at varying conversation depths. We noticed a high variance in the performance of different models when the higher conversation depths. Interestingly, the average performance peaked at turn 15, suggesting less dependence of the submissions on the conversation depth, as opposed to the passage ranking task.

\begin{table*}[]
    \centering
    \caption{Evaluating the Groundedness and Naturalness of the responses by GPT-4 model.}
    \begin{tabular}{llll}
    \toprule
        Group & Run     & Groundedness & Naturalness \\ 
    \midrule
        IRLab-Amsterdam & run-4-GPT-4 & 0.89 (65/8) & 4.0 \\ 
        InfoSense       & georgetown\_infosense\_ikat\_run\_3 & 0.67 (47/23) & 3.684 \\
        out-yahoo       & uot-yj\_run\_llmnoptkb & 0.67 (49/24) & 2.9178 \\
        IITD            & run\_automatic\_dense\_monot5 & 0.51 (37/36) &  2.808\\
    \bottomrule
    \end{tabular}
    \label{tab:groundedness}
\end{table*}

\subsection{Response Evaluation}
Table~\ref{tab:groundedness} lists the results, where we saw that the GPT-4-based model outperforms other models by a large margin. Given that the results are assessed also by GPT-4, then these results are likely to be somewhat biased towards GPT-4-based submissions. So we warn participants (and readers) not to take these results at face value, as (human) assessments are required to provide an independent evaluation.

\section{Conclusion}
The first TREC iKAT edition developed resources for studying personalized conversational information seeking and added to the community's understanding of the topic. As a successor of TREC CAsT, it made significant advances over CAsT, by focusing on more personalized and complex conversations that require advanced reasoning and leveraging of the personal knowledge graphs to provide relevant responses.  The PTKB statement ranking task provided a way for participants to leverage users' personal information into the conversation. 
In terms of passage ranking and Response generation, we observed that \RG approaches were outperformed by \GRG approaches (where all the top four automatic submissions were based on \GRG).
This signals a shift in strategy -- where first the LLM's internal knowledge is drawn upon by directly generating answers, and then the answers are grounded through the retrieval step, before the final response generation.

\section{Acknowledgments}
We are grateful to Jamie Callan's great help and patience in enabling us to provide the participants with a sample of the ClueWeb'22 collection.
We are thankful for Ian Soboroff's experience, patience, and persistence in running the assessment process. We are also thankful to Andrew Ramsay for his help throughout the project. Finally, we thank all our participants and annotators.
\bibliographystyle{splncs04}
\bibliography{bibliography}
\newpage

\end{document}